\documentstyle[epsf,12pt,epsfig]{article}

\newcommand{\al}{\mbox{$\alpha $}}

\newcommand{\be}{\begin{equation}}
\newcommand{\br}{\begin{eqnarray}}
\newcommand{\ee}{\end{equation}}
\newcommand{\er}{\end{eqnarray}}

\renewcommand{\theequation}{\thesection.\arabic{equation}}
\begin{document}
\title{
\hfill\parbox{4cm}{\normalsize IMSC/99/05/23 \\hep-th/9905219}\\ 
\vspace{2cm}
On the Role of Chaos in the 
AdS/CFT 
Connection}
\author{S. Kalyana Rama and B. Sathiapalan\\
{\em Institute of Mathematical Sciences}\\
{\em Taramani}\\
{\em Chennai 600113}\\
{\em INDIA }}
\maketitle
\begin{abstract}
The question of how infalling matter in a pure state forms \\ 
a Schwarzschild black hole that appears to be at non-zero temperature
is discussed in the context of the AdS/CFT connection. It is argued that
the phenomenon of self-thermalization in non-linear (chaotic) systems
can be invoked to explain how the boundary theory, initially at zero
temperature self thermalizes and acquires a finite temperature. Yang-Mills
theory is known to be chaotic (classically) and the imaginary part of the
gluon self-energy (damping rate of the gluon plasma)
is expected to give the Lyapunov exponent. We explain
how the imaginary part would arise in the corresponding supergravity
calculation due to absorption at the horizon of the black hole. 
\end{abstract}
\newpage
\section{Introduction}
  
Ever since the discovery by Bekenstein and Hawking that black holes
have an entropy \cite{bek}
and a temperature \cite{haw}
associated with them there has been a
puzzling question.
When infalling matter in a pure quantum state collapse into a black
hole, how does it become a mixed state, which a Schwarzschild black hole
with a finite Hawking temperature must needs be? One possible 
answer is that
quantum mechanics in the presence of gravity is not unitary. Another
is that the state is actually pure and only appears mixed to the outside
observer. In this paper we wish to dwell not on this question  but rather on a
 related simpler question: Why does the state
{\em appear} thermal to the outside observer
 (regardless of whether it is actually
pure or mixed)?  This second question in some sense can be phrased
in an operational manner since a thermal state (in the thermodynamic sense
of the word)
can be defined to be one
which for instance has a thermal (Maxwell-Boltzmann,
Bose-Einstein or Fermi-Dirac) distribution of particles. Or it can be defined
as a system that looks ergodic.

 Chaotic systems typically become ergodic
on a time scale given by the (inverse of the)
Lyapunov exponent. Thus it is possible for
a non-thermal state to become thermal. 
Therefore an answer to the above question could be that the system is chaotic.
Thus we could say that infalling matter
initially in a coherent non thermal state must self thermalize 
upon collapsing into a black hole. 
%due to its own dynamics. 

Of course this problem must be treated quantum mechanically. 
Quantum chaos is less well understood, but if we accept
Berry's conjecture as a definition of quantum chaos then it has been
shown that this thermalization takes place quantum mechanically also 
\cite{Sred}. In fact these ideas can be used to define  
statistical mechanics without invoking a God-given ``heat bath'' \cite{Sred}.

Of course it is rather difficult to study self-thermalization in a 
complicated situation such as black hole formation. The recent connection
between gravity and boundary conformal field theory \cite{JM,GKP,EW1,EW2}
 allows us to pose this question in terms
of flat space Yang-Mills theory. It is thus reassuring to know that
classical Yang-Mills exhibits chaos and self thermalization and that
the Lyapunov exponent is nothing but the damping rate of hot 
(quantum) gluon
plasma.  This has  been calculated for high temperature QCD 
\cite{Pis,Bpis,Biro} and seems to
agree with numerical results on chaos in Yang-Mills \cite{Muel}.

What is found \cite{Biro} is that the damping rate of a thermal
plasmon at rest is obtained in perturbation theory to be \cite{Pis}:
\[
\gamma _0 \approx 6.635 {N\over 24\pi}g^2 T 
\]
\[= 0.176 g^2T \, \, \, N=2 \]
\be 	\label{Ly}
=0.264 g^2T \, \, \, N=3 
\ee
 
The Lyapunov exponent on the other hand is found numerically 
 to be \cite{Biro,Muel}
\[ \lambda _0 \approx 0.34 g^2T \, \, (N=2)\]
\[ \approx 0.53 g^2 T  \, \, \, (N=3) \]
It is thus seen that $\lambda _0 \over \gamma _0$$=2$.
 (As explained in \cite{Biro} the factor of two can be traced to the
 way the two quantities are defined.) The numerical claculations
 \cite{Muel} are done on a lattice of approximately $(20)^3$ sites and
 for a range of values of the lattice spacing and $g$. It is found to
 be chaotic even for very small values of $g$ which measures the non-linearity.
At very high temperatures one might expect that classical results have
 some validity in the quantum regime also. This is also reflected in
 Berry's conjecture about the nature of quantum energy levels of a
 classically chaotic system - the assertions are about the higher
 energy levels. Therefor we assume that QCD with large enough energy density
exhibits thermalization.

If we further make the (reasonable) assumption that at high temperatures
Supersymmetric Yang-Mills is qualitatively similar to QCD 
(at high temperature)
 the puzzle is resolved: When we add some coherent energy density
to a Yang Mills vacuum the system thermalizes and becomes ergodic and then
one can associate a temperature $T$ related to the energy density in the 
usual way $E/V \approx T^4$.  In gravity this process corresponds to the 
formation of a black hole with a finite Hawking temperature. This is 
an answer to the (``simpler'') question posed above at least at a 
conceptual level.  It still remains to actually calculate this
exponent in supergravity.  This is similar to the calculation of the
glueball mass using supergravity.

The real part of the self energy in the confining phase, the glueball mass,
 in 3+1-QCD was calculated using supergravity \cite{COOJ,MKJM,Z}.
Similarly one can calculate the imaginary part of the self
energy in high temperature 3+1-Supersymmetric Yang-Mills. 
Let us understand the supergravity origin of the imaginary part. From simple
arguments one can show that the eigenvalues of a Schroedinger like
operator have imaginary parts when the boundary conditions are such that
the current going into a region is not equal to the current going out.
If we imagine say a (normalizable) solution describing a dilaton
outside the horizon of a black hole, with incoming flux equalling zero
at infinity, then, as the dilaton falls into the black hole the flux
at the horizon will be non zero.  Such a solution must have an imaginary
part. It is crucial here that the boundary condition just 
%inside 
outside the horizon
is that there are only ingoing waves. These are complex boundary conditions
and that is why the eigenvalues are complex. 
In Appendix A we give a very brief description of
a one-dimensional quantum mechanical problem that illustrates this point.
In the glueball mass calculation of \cite{EW2,COOJ,MKJM,Z},
the boundary condition was that the derivative vanish at the horizon. This 
ensures
that the eigenvalues are real. This boundary condition is correct for
an equilibrium situation.  Another way to see that this is correct is to
look at the imaginary time description of equlibrium statistical mechanics.
In this case the horizon is like the origin of a polar coordinate system and 
consisitency demands such a boundary condition. The non-equlibrium situation
on the  other hand does not require such a boundary condition. It requires the
complex boundary condition that we described above. 

One can then ask: If the complex boundary conditions corresponding
to non-equilibrium are due
to absorption by a black hole,    
how can there ever be equilibrium in such a situation i.e.
 how can a dilaton possibly not be absorbed by a black hole? 
The answer, obtainable only in a complete theory, 
is likely to be 
that Hawking radiation takes care of the emission and in equilibrium
 absorption and emission must be equal. The imaginary part is a measure
of how fast equilibrium is reached after a small perturbation disturbs the
system. It is equal to the difference between the two rates.
In equilibrium the imaginary part will be zero.

What does Hawking radiation correspond to in Yang-Mills? As it represents
a flow of energy from small to large radii, it must represent the flow
of energy from long wavelengths to short wavelengths of the Yang-Mills
fields, as the energy is equipartitioned in reaching thermal
equilibrium. 
This is due to the same non-linearity of the equations that
is also responsible for the phenomenon of chaos and self thermalization.
In thermal equlibrium energy is uniformly
distributed in all modes as demanded by
equipartition.

If the above idea is basically correct one can ask further questions, such as
what is the description within Yang-Mills of
the phenomenon of a particles falling into a black hole.  
The answer is known \cite{Triv}: 
A particle/wave packet in the bulk starting a distance 
$\delta \stackrel{>}{_\sim} 0$ away from the boundary 
corresponds in the boundary 
to a wave packet of width $\Delta X \simeq \delta$. 
As the particle in the bulk falls radially inwards, $\Delta X$ 
increases, by the usual IR/UV duality between 
the bulk and the boundary. When the particle reaches the horizon, 
$\Delta X$ reaches the thermal wavelength. 

Also, the speed of the infalling particle cannot exceed that of
light. This translates into an upperbound on the rate of wavepacket 
spreading in the boundary. This is causality in the bulk, and 
its manifestations in the boundary theory have been studied in 
\cite{sumit}. 

One can also
ask more refined questions such as what happens if the impact parameter
of the bulk particle is sufficiently large 
so that the particle never reaches
the horizon.  When the impact parameter is non zero, 
the particle has angular momentum 
which is conserved in the bulk, certainly so in the
supergravity approximation used to describe the bulk space time. 
Thus, as the particle falls inwards, its radial velocity decreases, 
its angular velocity increases. If the angular momentum is
sufficiently large, then the particle will not reach the horizon at
all.  This would correspond in the boundary theory to a wave packet
that never thermalizes.  This is somewhat counterintuitive.

Furthermore,
the bulk angular velocity corresponds to the boundary 
wave packet momentum $p_i$, 
in the boundary space time. 
The bulk angular 
momentum conservation should then translate into the boundary theory 
as the conservation of the ratio of the wave packet momentum and 
its size. Moreover, as the size increases (bulk particle's radial
location decreases), the wave packet momentum increases ! It is not 
obvious that the boundary theory (Yang-Mills theory at finite
temeprature) has such an unusual conserved quantity. 

It is in fact quite possible that there is no such conserved
quantity. AdS/CFT corresondence would then predict 
that the angular momentum of the bulk particle is not conserved. 
This is not such an outrageous prediction. Presumably what is happening is 
that 
the bulk particle is interacting with the Hawking particles emitted by 
the black holes, thereby loosing its angular momentum. 
In this situation any particle will fall into the black hole eventually
regardless of its initial angular momentum.
While the 
supergravity theory used for the bulk does not know of this 
 effect, the boundary theory somehow knows about this 
 and is predicting its effect. One would expect that if the
supergravity equations are corrected to include  self interaction
of dilatons, equivalently the back reaction of Hawking photons, 
this effect would be seen.

Regardless of whether one neglects the drag effect of the heat bath or
not, one thing is clear from the above.  If the wavepacket has a large
value of the boundary momentum $p_i$ it will take longer to
thermalize. It turns out that this is seen in numerical simulations
for Yang-Mills - the Lyapunov exponents are found to decrease with
$p_i$.  Thus the supergravity connection provides an intuitive
understanding of this fact.

If the ideas described above are correct one should be able to calculate
the Lyapunov exponent for Yang-Mills using the supergravity connection.
One should then compare with that obtained using finite temperature
methods in QCD. One should of course keep in mind
that the latter is computed in the perturbative 
regime whereas the supergravity calculation necessarily describes the strong
coupling region.  The inverse gluon two point function at finite
temperature has the form (ignoring the index structure):
\[
\Delta ^{-1}  (\omega ) = \omega ^2 - k^2
-\Pi (\omega ),
\]
 where $\Pi (\omega )$
contains the finite
temperature corrections to the tree level result.  This is calculated
in the imaginary time formalism where $\omega $ takes values
$2\pi i n T$.  We can analytically continue $\omega$ to the real axis
and locate the poles $\omega _{0} + i \gamma _{0}$. Clearly 
$\gamma _0 \approx {Im \Pi (\omega _0 )\over 2 \omega _0}$.  $\gamma _0$
in fact is the damping coefficient.  On the supergravity side rather
than calculate the propagator, we determine $\omega _0 + i\gamma _0$
directly by solving the dilaton equation.   
\footnote{The real-time method does not reproduce correctly the imaginary
part of the two
point function \cite{DJ,W}.  Nevertheless the {\em location} of the pole 
will be correct. This implies that the imaginary part of the self energy
function which determines the Lyapunov exponent is also correct.}

The computation proceeds as in \cite{EW2} except for two differences:
One is that we will assume a real time dependence and solve for the 
energy eigenvalue in terms of the momentum.  The second is that the
boundary condition will be different and will lead to a complex eigenvalue.
We will not be able to evaluate the eigenvalue analytically but we can
argue that it will be complex and also determine the overall scale. 
The numerical evaluation will be reported elsewhere. 

Finally we would like to make some comments on the issue of why
classical Lyapunov exponent has any significance in quantum Yang-Mills.
What we have calculated is the damping
coefficient of a hot (quantum) gluon plasma - 
this is a classical calculation in
supergravity, but it is fully quantum mechanical in Yang-Mills. 
It turns out that in perturbation theory the leading term is
independent of $\hbar$ (see eqn (\ref{Ly}))
 and is a classical result \cite{Biro,Pis}.
 It also turns out, as mentioned earlier, 
that this leading term is equal to the Lyapunov exponent of classical 
Yang-Mills. Thus we conclude from this that classical Yang-Mills is
chaotic. Now when we go to the quantum theory, presumably we must
retain the $\hbar$ corrections to the gluon plasma damping
coefficient. But the direct relevance of this quantity to quantum
chaos is not known. 
As mentioned earlier, it is generally believed 
that systems that are classically chaotic
are also quantum mechanically ``chaotic''. 
 If these are true
then quantum Yang-Mills (with sufficicient energy density) 
will also look thermal. It is this property of
classically chaotic systems that we are invoking to make statements
about the quantum theory.

To summarize, what we have calculated is the damping coefficient of
hot (quantum mechanical) gluon plasma. This is a physically
interesting quantity by itself. Moreover, its classical limit gives
the 
Lyapunov
exponent of classical Yang-Mills which, is thus chaotic. Qualitatively we
expect
that quantum Yang-Mills will also be chaotic and hence look
thermal. 
  
The role of the $\hbar$ corrections (that we have in effect computed)
to quantum chaos in quantum Yang-Mills is an open question. One can
reasonably conjecture that if one had an exact effective action for
hot (large N) QCD (i.e. after doing the functional integral)
its Lyapunov exponent would be this fully quantum mechanical damping
 coefficient that we have computed.

\section{Preliminaries}
\setcounter{equation}{0}

The metric describing an $AdS_5$ - Schwarzschild black hole space-time
is
\[
ds^2 = -({r^2 \over b^2} + 1 - {GM \over r^2})dt^2 + {dr^2 \over
({r^2 \over b^2} + 1 - {GM \over r^2})} + r^2 d \Omega ^2 .
\]
\be 	\label{AdS}
= -V(r)dt^2 + {dr^2 \over V(r)} + r^2 d\Omega ^2 .
\ee
 
The boundary is at $r \rightarrow \infty$.  So we let the boundary be at
$r=R $ and assume
$R$ is very large.  The boundary metric is then
\[
ds^2 = - {R^2 \over b^2} dt^2 + R^2 d\Omega ^2 .
\]
\be	\label{bound}
= -d\tau ^2 +   (dx^i)^2   
\ee

The period of $t$ is
\be
\beta _H = {4\pi b^2 r_+ \over 4r_+ ^2 + 2b^2 }
\ee

In the above we have used $r_+$ which is defined by:
\[
V(r)=\frac{1}{r^2 b^2}(r^2 - r_+ ^2 )(r^2 + r_- ^2 )
\]
Using $r_+ ^2 - r_- ^2 = - b^2$, we get another expression for
$\beta _H$
\be	\label{2.4}
\beta _H = {2\pi b^2 r_+ \over (r_+ ^2 + r_- ^2 )}
\ee

 $\beta _H $ defines a Hawking temperature $T_H = {1 \over \beta _H }$.  The boundary
temperature for the metric (\ref{bound}) is then given by
$T_B = \frac{b}{R} T_H$.

We can consider two limits.  If we let $GM << b^2$ we get $T_B = 
\frac{1}{2\pi R}(\frac{b}{\sqrt {GM}})$.  If we take the opposite limit
we get $T_B = \frac{1}{\pi R} 
\left( \frac{G M}{b^2} \right)^{\frac{1}{4}}$. 
Note that the latter limit is used in \cite{EW2} along with a further
change of coordinates ${r \over \rho}= ({GM \over b^2})^{1\over 4}$. 
In this limit the boundary becomes $R^3$ as against $S^3$ for finite 
$GM\over b^2$.

\section{Classical Trajectories and WKB Solution}

If we let $p_0 = E $, the energy, and $L$ be the angular momentum, these being
conserved, they serve to label the trajectories.  The mass shell
condition $g^{\mu \nu}p_{\mu} p_{\nu} =0$ gives us the equation
\be	\label{geod}
-{E^2\over V(r)} + \frac{1}{V(r)}(\frac{dr}{d\lambda})^2 + {L^2 \over r^2}
=0
\ee

Here $\lambda $ is an affine parameter that is defined by 
$p^r = {dr \over d\lambda }$.
Thus we get 
\be
(\frac{dr}{d\lambda})^2 = E^2 - V(r){L^2 \over r^2}=E^2 - U(r)
.\ee

The quantity $U(r)$ is analogous to a classical potential and is shown in
figure 1 in the case $({GM\over b^2}) \rightarrow \infty$.

\begin{figure}[htb]
\begin{center}
\mbox{\epsfig{file=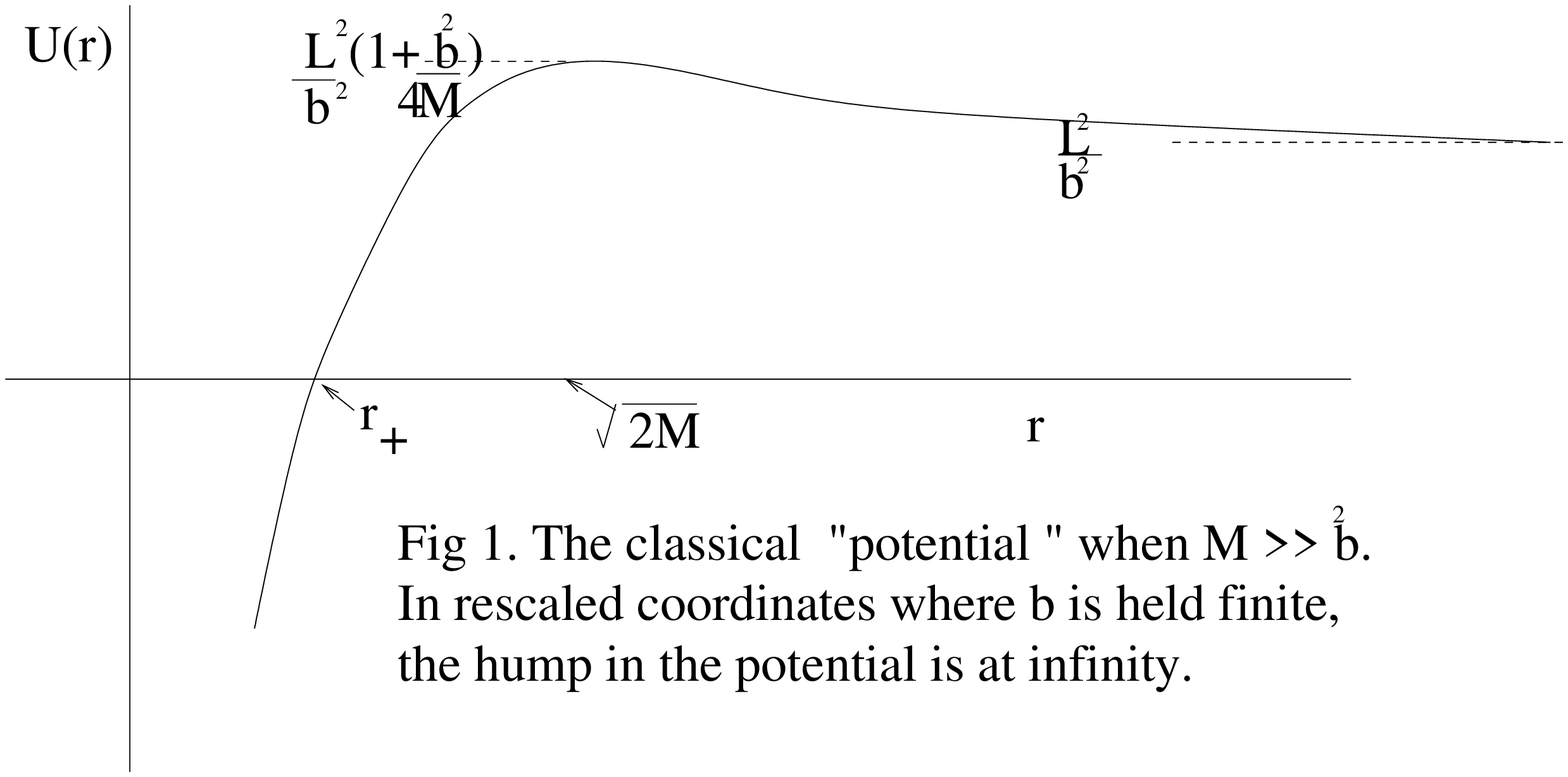,width= 6truecm,height=6truecm,angle=0}}
\end{center}
\end{figure}

The height of the potential is 
${L^2 \over b^2}(1+ \frac{b^2}{4 G M})$. Thus
whenever $E^2$ exceeds this the classical particle will fall into
the black hole.  Note also that in the limit taken in \cite{EW2}
the hump is at infinity.  So the region being dealt with is to the left
of the hump.  In this case all particles fall into the black hole. 

To get a WKB type solution we assume a solution of the form 
\be
f=e^{i{S(r)\over \hbar }}
\ee
 with $S=  S_0 + \hbar S_1 + ...$. 
On substituting in the wave equation (see (\ref{4.6}) below) one finds
\[
S_0 = \int {dr\over V(r)}\sqrt {E^2 - {L^2 V(r)\over r^2}}
\]
For large $r$ this becomes:
\[
S_0= -{b\over r}\sqrt {E^2 b^2 - L^2}
\]
\be
S_1 = {3i\over 2} ln \, r
\ee
The higher terms for large $r$ can be seen to involve higher powers of 
${r\over b \sqrt {E^2 b^2 - L^2}}$.  Thus it is a good approximation
for $r<< b$ and also
far from the turning point. The turning point $r^*$ is when $E^2 = U(r^*)$.
For $r<r^*$ the solution is damped, rather than oscillatory. 
Thus for $r<<r^*$ the field vanishes. 
The flux thus vanishes and the dilaton is not absorbed by the black hole. 
As shown in section 5 one can show in such cases that the eigenvalue
$E$ is real and therefore the Lyapunov exponent is zero. 
Thus we see as claimed in the introduction, that 
in this approximation when
 the impact parameter is larger than the black
hole radius the Lyapunov exponent is zero. 
\section{Dilaton Wave Equation}

In the above metric the wave equation for the dilaton is:
\be	\label{4.6}
{d\over dr}[r^3 V(r) {df\over dr}] + r^3 [-{s^2 \over V(r)} - 
{L^2 \over r^2}]f 
= 0.
\ee
Here the dilaton field has been written as:
\[
\Phi (t,r,\Omega) = e^{-st}f(r)Y_L(\Omega )
\]
where $L$ is the angular momentum and  $\Omega$ represents all the angles,
and $s = iE$ (with $E$ not necessarily real) specifies the time dependence. 
To be precise we can use the notation 
\be 	\label{4.6.5}
s\, = \,  i\, \omega + \gamma .
\ee  
We will assume
that $\omega , \gamma$ are greater than zero. Note that $\gamma$
 is the damping coefficient that
we want to determine.

We look for normalizable solutions.  This equation (\ref{4.6})
can also be written as

\be	\label{4.7}
{d^2f\over dr^2} + {[r^3 V(r)]'\over r^3 V(r)}{df\over dr} +
{1\over V(r)}[-{s^2\over V(r)} - {L^2 \over r^2}]f = 0.
\ee

For large $r$ as in \cite{EW2} the solutions either go to a constant
which is not normalizable or go as $1\over r^4$ which is normalizable.
Near $r=r_+$ the equation becomes:

\be	\label{4.8}
{d^2f\over dr^2} + \frac{1}{r-r_+}{df\over dr} + [{A\over (r-r_+ )^2}
+ {B\over (r- r_+ )}]f = 0 .
\ee
where
$A= -{s^2 b^4 r_+ ^2\over 4(r_+ ^2 + r_- ^2 )^2}$ 
 and $B = -{L^2 b^2 \over 2r_+ (r_+ ^2 + r_- ^2 )}$

Note that if we use (\ref{2.4}) we can rewrite 
\be	\label{4.8.5}
A= -{s^2 \beta _H ^2 \over 16\pi ^2}
\ee

The substitution $f = (r-r_+ )^\alpha \sum _{n=0} ^{n=\infty}a_n 
(r-r_+ )^n $ into (\ref{4.8}) gives the condition
$\alpha ^2 = -A$. Thus we can refer to $\pm \alpha $ as the two eigenvalues.
We should also note that if we set the energy parameter $s$ to zero
we get degenerate solutions.  If we let $y_{\pm \alpha}$ denote the
two solutions, then  in the degenerate limit $\alpha \rightarrow 0$,
 ${y_{+\alpha} - y_{-\alpha}\over \alpha}$ is 
the solution to consider and in fact goes as $ln (r-r_+ )$.

For $\al \ne 0$, a certain
linear combination of these two $y_{-\alpha}+\beta (\alpha , B)y_{+\alpha }$
is the one that is normalizable at infinity. For a given value of $\al$
and $B$ there is a number $\beta (\al , B )$ (that can be determined
numerically) such that $y_{\alpha} + \beta (\al, B ) y_{-\alpha }$ is
normalizable, i.e. the condition of normalizability fixes the functional
dependence of $\beta $
\footnote{Not to be confused with any inverse temperature.}
 on $\al $ and $B$.

In the case at hand  we will
see below that the boundary condition 
at the horizon requires that $\beta( \alpha , B) =0$.  
This is an eigenvalue condition and
gives us the energy eigenvalues as a function of $B$ or $L$. 
 In the limit that ${GM \over b^2} \rightarrow 
\infty $,  $L^2$ can be replaced upto some overall scale
by $k^2$ the three momentum squared \cite{EW2}.

\section{Boundary Conditions and Complexity of Eigenvalues}

We need to impose some boundary condition to fix the parameter $\beta$
of the previous section. The boundary can be chosen to be at any value
of $r$, and the boundary condition should reflect some physical
requirement. As explained in the introduction, in \cite{EW2} the 
boundary condition was that at $r=r^+$ the derivative of the dilaton 
field vanish. 
The justification for this is that in equilibrium 
one should consider a
Euclideanized time coordinate $\tau$
 as is common in finite temperature field theory.
Then the point $r=r^+$ is like the origin of a polar coordinate system 
in the $r- \tau $ plane.
If the field is to be well defined the radial derivative must vanish.

In our case we have a non-equilibrium situation. We have a real 
time coordinate.  So this argument is no longer applicable. The physical
 problem
we are trying to describe is that of a dilaton outside the black hole falling
into the black hole. A reasonable boundary condition is that just outside
the horizon the field must be purely ingoing.  There cannot be any outgoing
wave at the horizon.
  This is in accord with physical intuition 
that says once a particle reaches the horizon it cannot go out. 
We will thus impose this boundary condition at an infinitesimal distance
$\mu$
outside the horizon at $r=r_+ + \mu$. $\mu$ is just a regulator that will
be set to zero in the end.

Let us first show that the eigenvalues have to be complex.
We multiply by $f^*$ and integrate equation (\ref{4.6}) 
from $r=a$ to $r=b$ to get

\be
\int _a ^b f^* {d\over dr}[r^3 V(r) {df\over dr}]dr 
+ \int _a ^b r^3 [-{s^2 \over V(r)} - {L^2\over r^2}]f^*f dr =0.
\ee

Integrate by parts to get
\be
-\int _a ^b \Bigg\vert {df \over dr}\Bigg\vert ^2 r^3 dr 
+ [f^*{df\over dr} r^3 V(r)]\Bigg\vert _{r=a} ^{r=b}
+\int _a ^b r^3 [-{s^2 \over V(r)} - {L^2\over r^2}]f^*f dr =0.
\ee

Subtract from the LHS its complex conjugate.  The first term is
manifestly real and drops out of the difference.
 The second term minus its complex conjugate gives the difference in flux
at the points $r=a$ and $r=b$. In the last term only
$s^2$ can possibly be complex and so we finally obtain

\be	\label{4.17}
[f^*{df\over dr} r^3 V(r)- c.c.]\Bigg\vert _{r=a} ^{r=b} 
= \int _a ^b {r^3\over V(r)}[s^2 - (s^*)^2 ]f^* f dr
\ee
Let us take $a=\infty$ and $b= r_+ + \mu$.
If we have a normalizable function $f$ that vanishes at
infinity and boundary condition at $r_+  + \mu$ described
above, the LHS is clearly non-zero.  Thus $s \ne s^*$.
This proves that the eigenvalues are complex.

\section{Solutions}
The two solutions to (\ref{4.6}) are
\be
y_{\pm \alpha}= \sum _{n=0}^{n=\infty} a_n (r-r_+ )^{n \pm \alpha}
\ee

If one calculates the current ${1\over 2i}\sqrt g 
g^{rr}(\phi ^* {d \phi \over dr}- c.c.)$
one finds that near $r=r_+$ it is proportional to $\pm Im \, \,\al$.
 We will adopt the convention that $Im \, \, \al > 0$.
We want the current to be in the $-r$ direction for $r= r_+$.   
This means we pick
$y_{-\alpha}$ as the near horizon solution.
Thus if the solution that is normalizable at $\infty$ is 
\be	\label{5.11}
y(r)  = y_{-\alpha} + \beta \, y_{+\alpha} \; \; \;:\, 
 r\, > \, r_+ + \mu \; ,  
\ee  
then the boundary condition  requires that $\beta =0$.

This is an eigenvalue equation for $\al$ which we have seen has to
have complex solutions.  The actual value can be determined 
numerically. An important point to note here is that this equation
is independent of $\mu$.  Thus the Lyapunov exponent which is
proportional to the real part of $\al$ is independent of $\mu$.
As described in \cite{Biro} this exponent is equal to the gluon
plasma damping coefficient calculated in \cite{Pis,Bpis}. It was shown there
that  to leading order in $k^2$ (our $B$ or $L^2$) this quantity is a pure
number times $g^2 T$ {\em in perturbation theory}. In particular it
does not have any infrared divergence. This last fact is presumably
reflected in the $\mu$-independence mentioned above.
In fact  we find that it is $\mu$-independent even for finite
$B$. Thus the infrared divergences that were present in the perturbative
QCD calculations \cite{Pis,Bpis} are absent here. This is not surprising.
In this formalism (which corresponds to strong coupling QCD)
confinement and a mass gap are manifest. 
Therefore
one does not expect infrared divergences. 

Also if we use (\ref{4.8.5}) we see that $\al = {s\, \beta _H\over 4\pi}$.
Thus  $\alpha$ being some pure number
independent of ratios of dimensionful parameters, we automatically find that
$s$ and hence $\gamma$ (see (\ref{4.6.5})) the damping coefficient,
 is  some pure number times the temperature.
We do not expect to reproduce the $g$-dependence as we are in a strong
coupling regime and the calculation is non-perturbative.

We cannot explicitly solve $\beta (\al , B)=0$ because we do not know the
functional form of $\beta$.  This has to be done numerically 
\cite{wip} and will be reported elsewhere.
Nevertheless one can make the following observation. For a given
amount of energy, if the dilaton has higher $L$ it must take longer
to reach the horizon.  This is because higher $L$ would mean its radial
velocity is smaller and also the ``centrifugal barrier'' is higher.  Thus
thermalization should take longer. This in turn means that the Lyapunov
exponents should decrease with increasing $k^2$.  
Thus for instance one can expect a relation of the form
\[
\gamma (k^2  ) = \gamma _0 - c k^2  ; \, c\, > \, 0.
\]
 Thus there should be a critical $k^2$ at which $\gamma =0$ and there would
be no thermalization.
This corresponds to the impact parameter being larger than the 
black hole radius. But again as mentioned in the introduction, this
is due to the neglect of the drag due to the background Hawking radiation.
Presumably if this effect were included, $\gamma $, 
 while decreasing with increasing $k^2$,  would always be 
greater than zero.
This seems to agree
with numerical work on Yang-Mills done on the lattice \cite{Muel}.

\section{Conclusions}

We have discussed in this paper a  possible resolution of a puzzle
regarding black holes, which is simply stated: How does a pure state
evolve into a state that  appears to be at finite temperature?
This question is difficult to answer directly in a theory of gravity.
However using the AdS/CFT connection it can be posed as a problem in 
Yang-Mills. Namely if some energy is added in a coherent form 
say as an electric field excitation, at zero
temperature, how does the theory ``suddenly'' acquire a finite temperature?
That this must happen follows from the AdS/CFT connection. An answer
to this question is that the theory ``self-thermalizes''.  This is known
to be property of chaotic systems and Yang-Mills is classicaly known
to be chaotic.  This seems to answer the question qualitatively.

Note that there is no implication here that quantum mechanics
violates (or does not violate) unitarity. 
The statement is that the final state with the black
hole {\em appears} to be at finite temperature.  We are invoking the
argument \cite{Sred}
 that a pure state 
can have this property, in a chaotic system.  

A more
quantitative check is to calculate the Lyapunov exponents in  supergravity
and compare with Yang-Mills calculation. An agreement would not only
strengthen the AdS/CFT conjecture, it would also say that chaotic behaviour
is there and can explain many otherwise mysterious facts about black holes.  
We have set up the calculation and shown that the existence of a horizon
implies that there are complex eigenvalues. The dependence of the 
Lyapunov exponent on temperature is as expected from Yang-Mills theory 
on dimensional
grounds.  The actual calculation involves numerical work. The infrared
divergences that show up in perturbative QCD (for the non leading terms)
do not show up here.
  We also have an equation for the dependence of the
Lyapunaov exponent on $k^2$. But again solving it requires numerical work.
The qualitative fact that the exponent should decrease with increasing
$k$ can be argued on physical grounds from the supergravity viewpoint.
This is also in accord with \cite{Biro,Muel}.

Thus the role of chaos seems to be important. But more work remains to
be done to quantitatively support this idea.

\begin{center}
{\bf Acknowledgements}
\end{center}
We would like to thank N. D. Hari Dass, P. Majumdar and
S. Sinha for some useful discussions and also
R. Basu for bringing \cite{Biro} to our attention.

\appendix

\renewcommand{\thesection}{\Alph{section}}
\renewcommand{\theequation}{\thesection.\arabic{equation}}

\section{Appendix: Illustrating Complex Boundary Conditions in 
Quantum Mechanics}
\label{appena}
\setcounter{equation}{0}

We illustrate how complex boundary conditions lead to complex eigenvalues
in a simple one-dimensional quantum mechanical problem. Consider the
potential shown in fig 2.

\begin{figure}[htb]
\begin{center}
\mbox{\epsfig{file=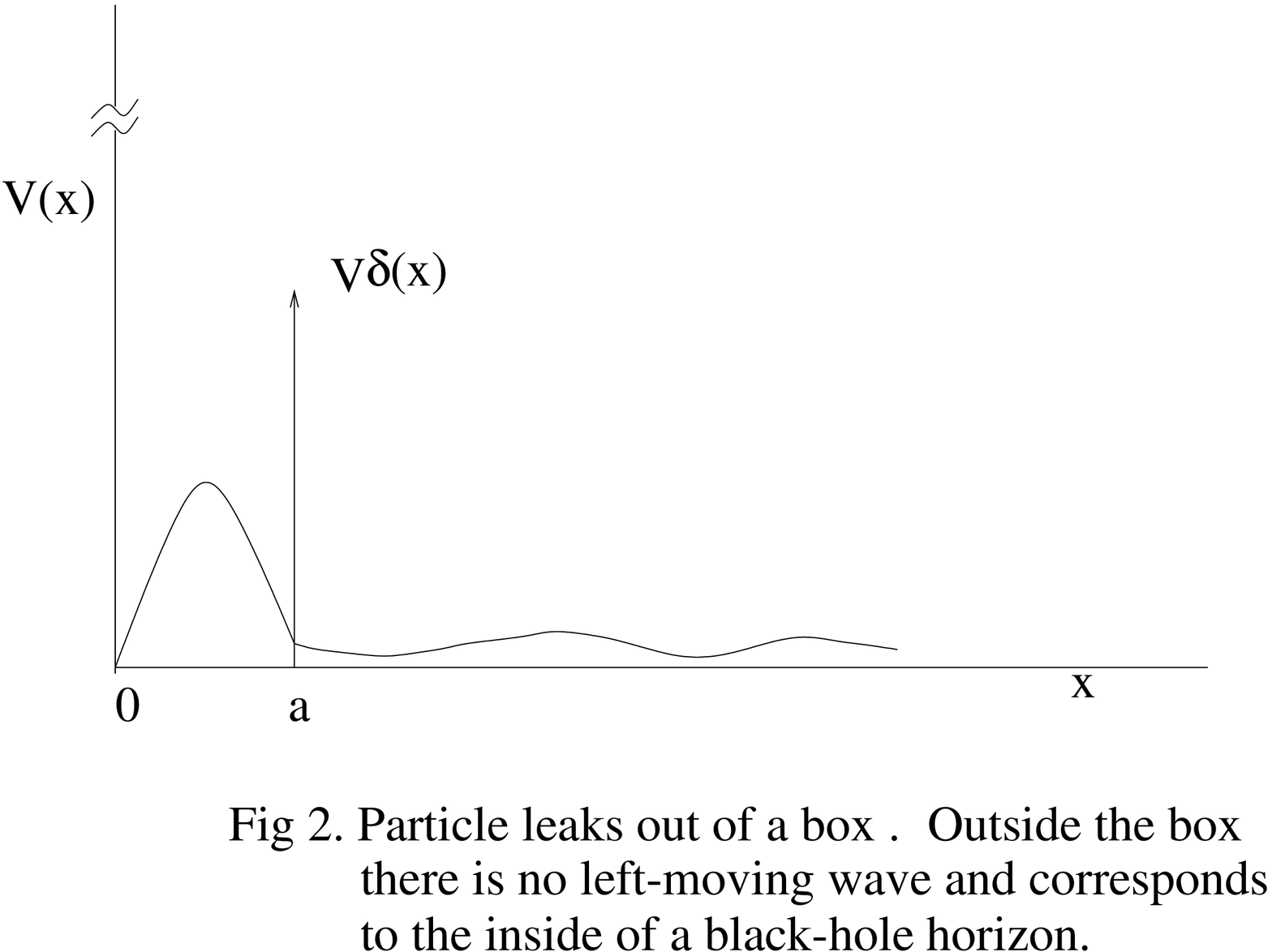,width= 6truecm,height=6truecm,angle=0}}
\end{center}
\end{figure}
 There is an infinite barrier on the left and a 
finite area barrier on the right,  which we represent by a $\delta $-function
of strength $V$ for simplicity. We have a particle in the box that,
for infinite $V$ is confined inside. But for finite, large $V$ will
slowly leak out.   
The wave function is given by 
\[
\psi = A(k) e ^{ikx} + B(k) e^{-ikx}  \, \, : \, \, x\, <\,a.   
\]
\be	\label{A1}
\psi = C(k) e ^{ikx}   \, \, : \, \, x\, >\,a.   
\ee

Note that outside the box we have assumed that the wave is purely right
moving. This is a complex boundary condition (at infinity). This is analogous
to the dilaton having only ingoing waves inside the horizon.

The other boundary conditions are
\be 	\label{A2}
\psi (0) =0
\ee
at $x=0$ and integrating Schroedinger's equation across the $\delta$-function
gives at $x=a$, in addition to continuity of $\psi$,
\be	\label{A3}
{d\psi\over dx}\Bigg\vert _{a-\epsilon} - {d\psi \over dx}\Bigg\vert
_{a+\epsilon}= -V\psi (a) 2m.
\ee

We look for ``stationary'' solutions with a time dependence
$e^{-iEt}$, with $E$ possibly complex. Thus
\be
{k^2\over 2m} = E
\ee

Imposing (\ref{A2}) gives $A=-B$, and imposing continuity at $x=a$
gives
\be
{C\over A} = 1-e^{-2ika}
\ee

Thus all constants $A,B,C$ are determined upto an overall constant that
can be fixed by normalization. However (\ref{A3}) still has to be
satisfied and this gives an eigenvalue equation for $k$  
\be	\label{A4}
2ik {e^{-2ika}\over 1-e^{-2ika}} = -V.
\ee

Let us set 
\[
k= p+i\, \mu
\]

When $V=\infty$ the solution clearly is $\mu =0$ and $p={n\pi \over a}$.
For large, finite $V$ we can do a perturbation expansion in powers of
$1\over V$ around this solution. We assume the wavenumber decreases a bit:
\[
p={\pi \over a(1+\epsilon )} ; \mu > 0.
\]
and $\epsilon ,\, \mu $ are small and non-zero.

Plugging into (\ref{A4}) we find to lowest order
\[
\epsilon = {1\over Va}
\]
\be
\mu = -{\pi ^2 \over V^2 a^3}
\ee
The fact that $\mu$ is negative means that the exponential is growing 
outside the box. In order that the wave function be normalizable, 
we have to assume that somewhere outside (but at a finite 
distance) there is a mechanism for absorbing the particle. This would be a 
physical justification for the boundary condition. In the case of the 
black hole this is what we assume happens, once the particle enters the 
interior of the black hole and reaches the singularity. 

Thus the real part of the energy is $p^2 - \mu ^2 \over 2m$ and
the imaginary part is the width $\Gamma = { p\, \mu \over m}
 \approx {\pi ^3 \over V^2 m^3 a^4}$

$\Gamma$ gives the rate of escape of the particle from the box (=flux)
and can be seen to be equal to $\mid C \mid ^2 \times \, velocity$, where
$velocity \, = {p\over m}$.

This example illustrates how energy becomes complex when complex
boundary conditions are used to describe leakage of flux. One can also
check that the equation analogous to (\ref{4.17}) for the Schroedinger
equation is satisfied for this problem.

\end{document}